\documentclass[prl,showpacs,nobibnotes,floatfix,superscriptaddress]{revtex4}

\usepackage{amsfonts,amsmath,graphicx,latexsym,color}

\begin{document}

\title{Origin of negative differential resistance in a strongly coupled
single molecule-metal junction device
}

\author{Ranjit Pati\footnote{e-mail:patir@mtu.edu}}
\affiliation{Department of Physics, Michigan Technological University, Houghton MI 49931, USA}

\author{Mike McClain}
\affiliation{Department of Physics, Michigan Technological University, Houghton MI 49931, USA}

\author{Anirban Bandyopadhyay}
\affiliation{International Center for Young Scientists, National Institute of Materials Science, 1-1 Namiki, Tsukuba, Ibaraki 305-0044, JAPAN}

\begin{abstract}
{\footnotesize 
A new mechanism is proposed to explain the origin of negative differential
resistance (NDR) in a strongly coupled single molecule-metal junction.
A first-principles quantum transport calculation in a Fe-terpyridine linker
molecule sandwiched between a pair of gold electrodes is presented.
Upon increasing applied bias, it is found that
a new $phase$ in the broken symmetry 
wavefunction of the molecule
emerges from the mixing of occupied and
unoccupied molecular orbital. As a consequence, 
a non-linear change in the coupling between molecule and lead is evolved
resulting 
to NDR. This model can be
used to explain NDR in
other class of metal-molecule junction device.
}
\end{abstract}


\maketitle 

\newpage

The controlled transport of electrons in metal-molecule junction has been
an active field of research for the last decade \cite{1}, with an aim to find
a possible solution to the miniaturization impasse that semiconductor
industry is
facing currently. Researchers have already demonstrated conduction, 
rectification, and switching in metal-molecule junction devices \cite{2,3,4}.
Among all,
the demonstration of a single-molecule switch with a
NDR feature \cite{2} has drawn considerable
attention
in recent years.
The NDR feature is described by a
steady increase followed by a decrease in current ($I$)
with the increase in applied
bias ($V$).
Since its realization \cite{2},
various groups have been working
on this problem to unravel the $true$ mechanism of NDR in molecular junction,
understanding of which would revolutionize the field of molecular
electronics \cite{1}.
Different
mechanisms have been proposed to explain the observed NDR \cite{2,5,6,7}
in metal-molecule junctions. For example, in
{\em weakly
coupled} junction, it is argued \cite{5,8,9}
that the narrow density of states (DOS) feature
of the tip apex atom is responsible for the NDR.  But in contrast, very 
recently, it is
proposed \cite{6} that the local orbital symmetry matching
between the tip
and the molecule is the cause of the observed NDR. In another example 
\cite{2}, 
the bias induced
charging and conformational change of the molecule is suggested as
the viable mechanism
for inducing NDR. However, it should be noted that controversies still remain
regarding
whether the conformational change arising from the rotaion
of the ligand group \cite{10}
or from the rotation of the molecular group \cite{11,12,13} within the
molecular
backbone is responsible for the observed NDR. 
Another possible model for NDR based on polaron
formation was also proposed very recently \cite{14}. Based on a simple
mean-field theory it is suggested \cite{14}
that NDR effect can be possible if the bias
induced polaron formation would shift the energy level away from the window
between the chemical potentials of the leads.

It can be seen from the above review that none of the proposed mechanisms 
thus far have stressed the
importance of bias induced coupling change between molecule and lead.
Bias induced
relaxation of the molecular eigenstates will affect the coupling between
molecule and lead, which would have a measurable effect
in altering the $I-V$ feature.
Furthermore, it is still controversial whether the NDR is intrinsic to the 
molecule or junction dependent. 
Here, we propose a {\em new consistent and unified model} to explain the
origin of NDR
in a strongly coupled metal-molecule junction.  
It is found 
that a new $phase$ in the broken symmetry wavefunction of the molecule
arising from the mixing of occupied and unoccipied molecular orbital
upon increasing applied bias
leading to a non-linear change in the coupling between molecule
and lead is responsible for the observed NDR.
Our calculation further reveals, even if we have a symmetric molecule-lead
junction at zero bias, the asymmetric channel coupling at the
junction upon applied
bias originating from the phase reversal of the molecular eigenstate 
can produce asymmetric I-V characteristic. 

We have used a self-consistent {\em many body}
approach to investigate the quantum transport properties of a Fe-terpyridine
linker molecule (FETP) sandwiched between two gold electrodes
(Fig. 1). The choice behind the selection of FETP molecule is prompted by the
\begin{figure}
\centering
\includegraphics [scale=0.5]{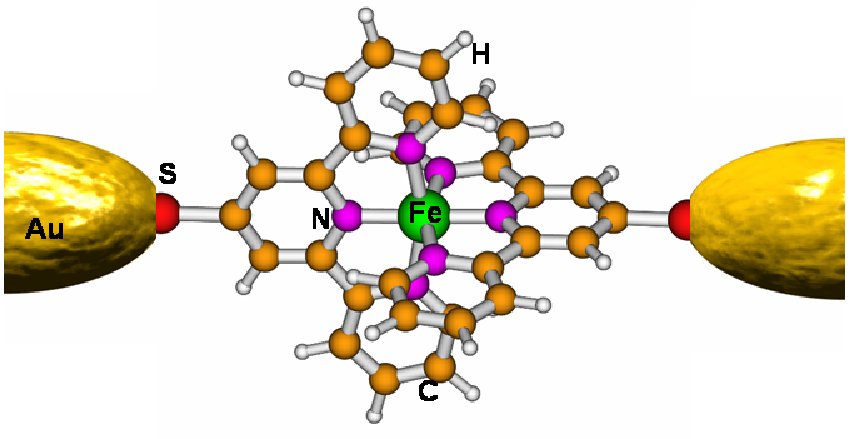}
\caption{(color online) Schematic of the strongly coupled single molecule
(FETP)-gold junction.}
\end{figure}
recent interest in organo metallic molecule \cite{15} in molecular electronics. 
Furthermore, very recently it has been shown organo-metallic molecule
exhibiting NDR behavior \cite{5}.
The electric field effect for each applied bias \cite{16}
is explicitly included in our calculation within a {\em many body} framework.
The non equilibrium Green's function (NEGF)
 approach is used to calculate the quantum
transport. Our calculation reveals strong-field induced asymmetry and NDR
features with peak ($I_{p}$) to valley ($I_{v}$) current ratio (PVR) of 2.7
at both
positive and negative bias. Even though the magnitude of calculated current
differs between the positive and negative bias range,
the $\frac{I_{p}}{I_{v}}$ remains about the same for both positive and negative
bias.

In our calculation, we have used a real space approach in which the {\em 
many body}
wavefunction is expanded in terms of finite set of atomic Gaussian orbital 
\cite{17}. 
 This allows us to partition the {\em open} molecular device structure into two
 regions of interest. The first is the {\em true} device region, which includes
 optimized FETP sandwiched between two 3 atoms gold clusters taken from
 Au (111) surface with the terminal S-atom at the three fold
 hollow fcc site\cite{18} of the gold with S-Au distance 2.45 $\AA$.
 The second
 part is essentially the semi-infinite contact region, which is essentially
 assumed to be unperturbed by the molecular adsorption. As exact exchange with
 dynamical correlation plays a key role in determining accurate energy spectra
 in density functional theory, and consequently the $I-V$ feature, we have used
 a $posteriori$ Becke's three
 parameter hybrid density functional approach (B3LYP) \cite{17,19} for our
 calculation.
 Los Alamos double zeta
 effective core potential basis set (LANL2DZ) \cite{17}, which includes the
 scalar
 relativistic
 effect, is used for the calculation of the
 non equilibrium electronic structure.
 The non equilibrium Hamiltonian, $H(\varepsilon)$  of the true device region
 is calculated as \cite{20}, $H(\varepsilon)=H(0)+\vec{\varepsilon}.
{\sum_{i}\vec{r}(i)}$,
where $H (0)$ is the equilibrium {\em many body} Hamiltonian; $\vec{\varepsilon}$ is the
 applied dipole 
electric field, and $\vec{r}(i)$ is the coordinate of the electron
 $i$.
 This approach allows us to explicitly obtain the $true$
non-equilibrium energy spectra including Stark shift. It should be noted
 that the convergence
thresholds for energy, maximum and root mean square electron density are set
 at $10^{-6}$ a. u., $10^{-6}$ a. u. and $10^{-8}$ a. u. respectively to
 ensure $tight$ $convergence$ during self-consistent calculation. Subsequently,
 we constructed the NEGF $G$ as:
$G(\varepsilon)=
(E\times I-H_{Mol}(\varepsilon)-\Sigma_{l}(\varepsilon)-\Sigma_{r}
(\varepsilon))^{-1}$,
where $H_{Mol}(\varepsilon)$ is the orthogonalized  field dependent Kohn-Sham
(KS) molecular
Hamiltonian  matrix obtained by suitable partitioning of $H(\varepsilon)$;
$E$ is the injection energy of
the tunneling electron; $I$ is the identity matrix. $\Sigma_{l,r}(\varepsilon)$
are the self-energy
functions \cite{21,22} calculated from the $bias$ $dependent$ coupling
matrices and the Green's function of the Au lead. The latter is approximated
from the 6s-band density of states at the
Fermi energy
of bulk Au \cite{23} and kept fixed for all bias points.
The current in the molecular junction is calculated as \cite{14,15}:
$
I=\frac{2e}{h}\int_{\mu_{1}}^{\mu_{2}} T(E,V)\times
 [f(E,\mu_{2})-f(E,\mu_{1})]
\times dE,
$
where $\mu_{1,2}=E_{f}\mp
eV/2$; $E_{f}$ is the equilibrium Fermi energy (-4.69 eV). $f$ is the fermi
 distribution function. $T(E,V)$ is the bias
dependent transmission function \cite{15} obtained from non-equilibrium 
$G$ and $\Sigma$s.
\begin{figure}
\centering
\includegraphics{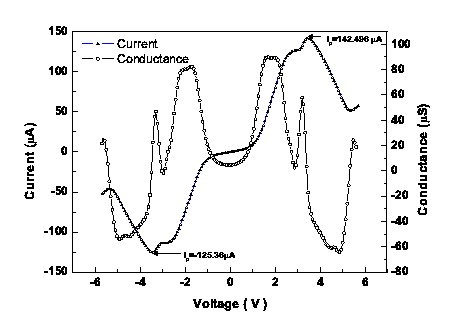}
\caption{Calculated current and conductance as a function of applied bias.
$I_{p}$ refers to the peak value for current.}
\end{figure}

First we comment on our calculated $I-V$
characteristics presented in Fig. 2.  The $I-V$ characteristic is asymmetric
 and non-linear.  The current is symmetric for voltages up to +0.4 V and -0.4 V.
 As the bias increases the asymmetry takes over and the magnitude of current
 increases till it reaches a peak current of -125.36 $\mu A$ at -3.417 $V$,
 and
 142.496 $\mu A$ at 3.484 $V$ respectively. Increasing the applied bias
 further on
 both the positive and negative bias range, the magnitude of current decreases
 and reaches a valley current of -45.78 $\mu A$ at -5.36 $V$, and
 51.69 $\mu A$ at 5.36 $V$
 respectively-$revealing$ $ a$ $ clear$  NDR $ pattern$. The peaks to
 valley ratio (PVR)
 in currents are found to be 2.74 and 2.76 respectively on the negative and
 positive bias range.  To corroborate the NDR patterns in I-V, we calculated
 the conductance, $dI/dV$, which is plotted in  Fig. 2.
 The negative values of the conductance in the positive and negative bias
 range clearly confirm the two peaks. In addition, we found another small
 negative value for conductance at -2.91 $V$.
 Examining the I-V curve, we found from -2.8 $V$ to -3.0 $V$, the
 current
 is almost constant ($blockade$ $effect$) with a very small
 0.1 $\mu A$ spike at -2.88 $V$. 
 Since the bias range considered in
 this study is quite high, we included the incoherent scattering effect (ISE)
 due to electron-phonon $(e-p)$ coupling self consistently within
 B$\ddot{u}$ttiker's approach \cite{17,24,25} and recalculated the current at
 -2.88 $V$
 as well as
 for the bias points around it. We found the small spike of
0.1 $\mu A$ at -2.88 $V$ disappear. Inclusion of $e-p$ coupling at bias
 points -3.417 $V$
 and -5.36 $V$ gives $I_{p}$ and $I_{v}$ as -120.17 $\mu A$ and -52.21 $\mu A$
 with PVR of 2.30. The PVR value is found to be 2.34 for the positive bias
 range with the inclusion of ISE.  Inclusion of ISE is found to reduce the
 current, but do not shift the peak or the valley position of the current.
 It is worthwhile to note that for the calculation of $I-V$ in which the
 explicit
 field
 dependent term is not included in $H(\varepsilon)$,
 the result do not reveal asymmetry or NDR
 effect, 
suggesting the importance of self-consistent electric field dependent
 screening in inducing NDR. 
\begin{figure}
\centering
\includegraphics{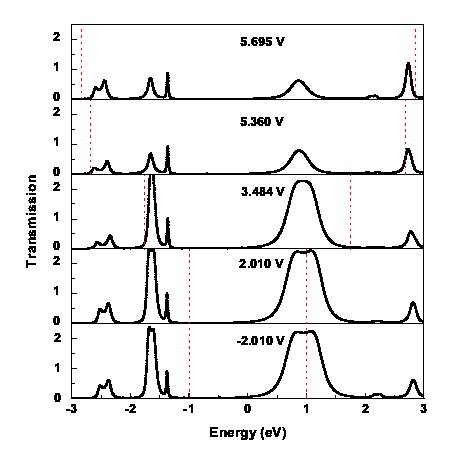}
\caption{Bias dependent transmission as a function of injection energy $E$.
Fermi energy is set to zero in the energy scale; Dotted lines in each panel
represents the chemical potential window.}
\end{figure}

 To understand the origin of asymmetry in current
 and NDR, we have calculated the bias
 dependent transmission function (Fig. 3) as a function of injection energy.
 For brevity, we have considered only five bias points.  The higher current
 in the positive bias range (80.42 $\mu$A at 2.01 V) as compared to that for
 negative bias (-74.69 $\mu$A at -2.01 V) can be
 understood by comparing the transmission function for -2.01 $V$ and
 2.01 $V$.
The higher transmission around $\sim$ 1 $eV$ 
for 2.01 $V$ (within the chemical potential window) as compared to that in
-2.01 $V$
explains the higher current at the bias point 2.01 $V$ than -2.01 $V$.
Analysis of eigenvalues of the $H_{Mol}(\varepsilon)$, and the 
DOS calculated from $G(\varepsilon)$, suggest that the lowest
unoccupied molecular orbital (LUMO) and LUMO+1 (Fig.4) contributes to the 
current at 2.01 $V$ and -2.01 $V$. No change in eigenvalue spectrum of 
$H_{Mol}(\varepsilon)$ is found between 2.01 $V$ and -2.01 $V$, suggesting
the importance of channel coupling with the lead in inducing the asymmetry.
Increasing the bias from 2.01 $V$ to 3.484 $V$ (peak position),
we found a small decrease in the contribution of LUMO and LUMO+1 as the
transmission peak height at $\sim 1 eV$ shows a small decrease.
But, as the chemical
potential window increases, the highest occupied molecular orbital (HOMO)
and HOMO-1, and HOMO-2 channels (Fig. 4) contribute to the transmission at
$\sim$ -1.4 $eV$ and 
$\sim$ -1.6 $eV$, resulting to a higher net
transmission and current.
Increasing the bias from 3.484 V to 5.36 V (valley position), we found the 
contribution of LUMO and LUMO+1 to transmission decreases significantly,
so also the
contribution from HOMO and
HOMO-1, and HOMO-2. 
This explains why we see less current at 5.36 $V$ compared
to that at 3.484 $V$, despite the contribution of additional channels HOMO-3 ($\sim$ -2.3 $eV$)
and HOMO-4 ($\sim$ -2.5 $eV$) to transmission due to the increase 
in chemical potential window.
Increasing the bias further to 5.695 $V$,
we found the contribution of
LUMO and LUMO+1 decreases by a very small amount, but the contribution of
additional channels LUMO+2, LUMO+3 ($\sim$ 2.1 eV),
and LUMO+4 ($\sim$ 2.8 $eV$) to transmission increases. Furthermore, the
contribution to transmission from HOMO-3 and HOMO-4 also increases, leading to a higher
net transmission and current at 5.695 $V$ compared to that at 5.36 $V$. 
Another very interesting feature is
noted from the comparison of the transmission in the three upper panels of
Fig. 3. As the bias increases, the contribution to transmission from the
orbital near the
Fermi energy decreases,
but the
contribution from the orbital away from the Fermi energy 
increases. This clearly suggests
that the frontier
orbitals and the orbitals away from the Fermi 
energy respond differently to the applied bias. 
\begin{figure}
\centering
\includegraphics [scale=0.5]{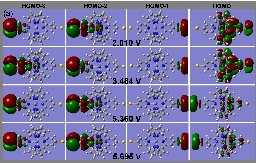}
\includegraphics [scale=0.5]{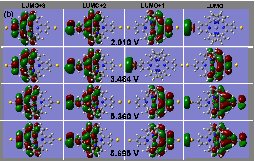}
\caption{(color online) (a) Occupied molecular orbital under applied bias; from
right to left (column wise), the orbitals are HOMO, HOMO-1, HOMO-2, and HOMO-3.
(b) Unoccupied molecular orbitals under applied bias; from right to left (column wise), the orbitals are LUMO, LUMO+1, LUMO+2, and LUMO+3. Bias changes row
 wise; Notation: dark gray (red) corresponds to negative lobe, and light
gray (green) corresponds to the positive lobe.}
\end{figure}

Thus the question arises- what is the cause for this non-monotonic feature
 in the
net
transmission which gives rise to NDR effect? Does NDR have a molecular
origin or is it junction dependent?
To answer these subtle 
questions, 
the eigenvalue spectrum of $H_{Mol}(\varepsilon)$ is analyzed near the
Fermi energy.
Though some small shift in energy levels bringing LUMO and LUMO+1 energy level
closer are found with the increase of bias from 3.484 $V$ to 5.36 $V$,
no shift in
energy levels away from the window of
chemical potential is observed,
which would explain the NDR.
But comparing the eigenvalue spectrum of $H_{Mol}(\varepsilon)$ and the
DOS obtained 
from $G(\varepsilon)$, a strong bias dependent energy level
broadening effect due to the coupling of the molecule
with the
lead is observed. Thus the question again arises-
why does the broadening effect change significantly from peak to
valley position? 
Broadening effect depends strongly on the coupling of molecular eigenstates 
with the lead. 
Thus we analyze the molecular orbital's response to the electric field (Fig.4).
First, between 2.01 $V$ and -2.01 $V$ only a phase reversal of the molecular
eigenstate is found. Second, 
as the bias increases (Fig.4) from 3.484 $V$ to 5.36 $V$,
a new phase in the broken symmetry wavefunction of
the molecular \cite{22} is
emerged because of the strong
mixing of the HOMO and LUMO+1 at 3.484 $V$ to give LUMO at 5.36 $V$.
This significant change is expected to have a strong effect on the coupling.
To confirm this hypothesis,
we recalculated the current
at 5.36 $V$ using coupling matrices extracted
at 3.484 $V$. The calculated current is found to be 62.63 $\mu A$, which is
higher than
the original
valley current. Similar calculation 
at 5.695 $V$  
using the bias dependent coupling matrices extracted from the self-consistent
calculation at 5.360 $V$
(valley position) gives a lower current of 50.05 $\mu A$
than the original valley current, 
suggesting the coupling is weaker at the valley position than at 3.484 $V$ and
5.695 $V$.
Thus, unambiguously, we have confirmed that
the non-linear coupling change arising from a bias dependent transition in  
broken symmetry phase
of the molecular eigenstate 
is responsible for the NDR.

In summary, using an exhaustive first-principles approach, in which the
electric field effect is explicitly included within a {\em many body}
framework, 
we have identified the origin of NDR in a strongly coupled single-molecular
junction. Our calculation in a FETP molecule-metal junction reveals asymmetry
and strong NDR feature in the $I-V$, with high PVR of $\sim 2.7$ at both
positive and negative bias. The origin of asymmetry in 
current between the positive and negative bias range is ascribed
to the asymmetric channel coupling at the junction 
arising from the phase reversal of the molecular eigenstate.
The high PVR suggests that this switching
device could potentially be used as an active component in a new generation
molecular electronics circuit.  Most important, we found that the
bias dependent
transition in the 
broken symmetry phase of the molecular wavefunction leading to 
a non-linear change in the coupling
between molecule and lead is the root cause for NDR. This mechanism
can be
used to explain NDR in 
other class of metal-molecule junction device. 

This work is supported by NSF through Grants No. ECCS-0617353,0643420.


\begin{thebibliography}{25}

\bibitem{1} A. Nitzan and M. A. Ratner, Science {\bf 300}, 1384 (2003). 

\bibitem{2} J. Chen, M. A. Reed, A. M. Rawlett, J. M. Tour, Science {\bf 286},
1550 (1999).

\bibitem{3} M. A. Reed {\em et al.}, Science {\bf 278}, 252 (1997).

\bibitem{4} N. J. Tao, Nature Nanotechnology {\bf 1}, 173 (2006). 

\bibitem{5} Y. Xue, {\em et al.},
Phys. Rev. B {\bf 59}, 7852(R), 1999.

\bibitem{6} L. Chan {\em et al.}, Phys. Rev. Lett. {\bf 99}, Art. No. 146803
(2007).

\bibitem{7} J. He, {\em et al.},
J. Am. Chem. Soc. {\bf 128}, 14828 (2006).

\bibitem{8} N. D. Lang, Phys. Rev. B. {\bf 55}, 9364 (1997).

\bibitem{9} I. -W. Lyo and Ph. Avouris, Science {\bf 245}, 1369 (1989).

\bibitem{10} M. Di Ventra, S.-G. Kim, S. T. Pantelides, and N. D. Lang, Phys. Rev. Lett. {\bf 86}, 288 (2001).

\bibitem{11} J. M. Seminario, A. G. Zacarias, and J. M. Tour, J. Am. Chem. Soc.
{\bf 122}, 3015 (2000).

\bibitem{12} J. Cornil, Y. Karzazi, and J. L. Bredas, J. Am. Chem. Soc.
{\bf 124}, 3516 (2002).

 \bibitem{13} J. Taylor, M. Brandbyge, and K. Stokbro, Phys. Rev. B {\bf 68},
121101(R) (2003).

\bibitem{14} M. Galperin, M. A. Ratner and A. Nitzan, Nano Lett. {\bf 5},
125 (2005).

\bibitem{15} J. Park, {\em et al.}, Nature {\bf 417}, 723 (2002).

\bibitem{16} We have 
applied dipole electric field ($\varepsilon$) in the increment of 0.0001 a.u.
along the positive and negative molecular axis connecting the two electrode. From the optimized length ($L$) of the molecule (13.07 $\AA$) between the gold
 electrode, we converted the electric field into potential difference ($V$)
using $\varepsilon=V/L$, which gives the corresponding increment in bias to
0.067 $V$.

\bibitem{17} Gaussian 03, Gaussian Inc., Pittsburgh, PA, 2003.

\bibitem{18} J. Zhou and F. Hagelberg, Phys. Rev. Lett. {\bf 97}, Art. No. 045505 (2006).

\bibitem{19} R. G. Parr and W. Yang, {\em Density-Functional Theory of Atoms and Molecules} (Oxford Science, Oxford, 1994).

\bibitem{20} A. Szabo and N. S. Ostlund, {\em Modern Quantum Chemistry} (Dover Pub. Inc., New York, 1996).

\bibitem{21} S. Datta, {\em Electron Transport in Mesoscopic Systems} (Cambridge University Press, Cambridge, 1997).

\bibitem{22} R. Pati, L. Senapati, P. M. Ajayan, and S. K. Nayak, Phys. Rev. B
{\bf 68}, 100407(R) (2003).

\bibitem{23} Periodic calculation is performed using Vienna ab initio Simulation Package (Technische Universit$\ddot{a}$t Wien 1999) with $32\times 32
 \times 32$ $k$-point mesh within the Monkhorst-Pack scheme; G. Kresse and J. Furthm$\ddot{u}$ller, Phys. Rev. B {\bf 54}, 11169 (1996). The value of the 6s-band DOS
is found to be 0.035 states$/$eV$\times$spin.

\bibitem{24} W. Tian, {\em et al.},
J. Chem. Phys. {\bf 109}, 2874 (1998).

\bibitem{25} To include the vibronic coupling effect, an additional self-energy
 function ($\Sigma_{p}(\varepsilon)$=-0.001 $\times$ 
$G(\varepsilon)$) is included  in $G(\varepsilon)$ and is determined
 self-consistently. The transmission for the three probe system is calculated using
$T(E,V)=T_{lr}+\frac{T_{pr}T_{pl}}{T_{pr}+T_{pl}}$ as described in Ref. 21 and 24.

\end{thebibliography}
\end{document}